\documentclass[aps,pra,floatfix,twocolumn,amsmath,amssymb,showpacs,10pt]{revtex4-1}

\usepackage{dcolumn}
\usepackage{graphicx}
\usepackage{pst-all,pstricks-add}

\newcommand{\ket}[1]{\lvert #1 \rangle}           
\newcommand{\bra}[1]{\langle #1 \lvert}           
\newcommand{\expv}[1]{\langle #1 \rangle}           

\newcommand{\rhoop}{\hat{\rho}}

\newcommand{\rhoi}{\hat{\rho}_\mathrm{i}}

\newcommand{\rhoo}{\hat{\rho}_\mathrm{0}}
\newcommand{\rhoone}{\hat{\rho}_\mathrm{1}}
\newcommand{\rhotwo}{\hat{\rho}_\mathrm{2}}
\newcommand{\rhof}{\hat{\rho}_\mathrm{f}}
\newcommand{\rhoprep}{\hat{\rho}_\textrm{i}}
\newcommand{\rhoprepdiag}{\hat{\rho}_\textrm{prep diag}}
\newcommand{\rhofdiag}{\hat{\rho}_\textrm{f diag}}
\newcommand{\rhox}{\hat{\rho}_\textrm{$x$}}

\newcommand{\rhofx}{\hat{\rho}_\textrm{f $x$}}

\newcommand{\diag}[1]{d_{#1}}
\newcommand{\diagf}[1]{d_{\textrm{f}\; #1}}
\newcommand{\dotdiagf}[1]{\dot{d}_{\textrm{f}\; #1}}
\newcommand{\cdiag}[1]{c_{#1}}
\newcommand{\cdiagf}[1]{c_{\textrm{f}\; #1}}
\newcommand{\Hs}{H_\textrm{s}(\lambda)}
\newcommand{\Hseq}{H_\textrm{seq}(\lambda)}
\newcommand{\Hseqper}{H_\textrm{seq/channel}(\lambda)}
\newcommand{\Hcorr}{H_\textrm{corr}(\lambda)}
\newcommand{\Hcorrper}{H_\textrm{corr/channel}(\lambda)}
\newcommand{\Hspure}{H_\textrm{s pure}(\lambda)}
\newcommand{\Hoptpure}{H_\textrm{pure opt}(\lambda)}

\newcommand{\Gseq}{G_\textrm{seq}(\lambda)}
\newcommand{\Gcorr}{G_\textrm{corr}(\lambda)}
\newcommand{\Gcorrseq}{G_\textrm{corr/seq}(\lambda)}
\newcommand{\score}{\hat{L}}
\newcommand{\sigmax}{\hat{\sigma}_x}
\newcommand{\sigmay}{\hat{\sigma}_y}
\newcommand{\sigmaz}{\hat{\sigma}_z}

\newcommand{\iop}{\hat{I}}
\newcommand{\xop}{\hat{X}}
\newcommand{\lambdaest}{\lambda_\mathrm{est}}
\newcommand{\lambdacutoffseq}{\lambda_\mathrm{cutoff}}
\newcommand{\mopt}{m_\mathrm{opt}}
\newcommand{\estimate}{\tilde{\lambda}}
\newcommand{\myvector}[1]{\boldsymbol{\mathrm{#1}}}

\newcommand{\uprep}{\hat{U}_\mathrm{prep}}
 
\DeclareMathOperator{\Trace}{Tr}
\DeclareMathOperator{\variance}{var}

\begin{document}

\author{David Collins}
\affiliation{Department of Physical and Environmental Sciences, Colorado Mesa University, Grand Junction, CO 81501}
\email{dacollin@coloradomesa.edu}
\thanks{Author to whom correspondence should be addressed.}

\author{Jaimie Stephens}
\affiliation{Department of Physical and Environmental Sciences, Colorado Mesa University, Grand Junction, CO 81501}

\title{Depolarizing channel parameter estimation using noisy initial states}

\begin{abstract}
 We consider estimating the parameter associated with the qubit depolarizing channel when the available initial states that might be employed are mixed. We use quantum Fisher information as a measure of the accuracy of estimation to compare protocols which use collections of qubits in product states to one in which the qubits are in a correlated state. We show that, for certain parameter values and initial states, the correlated state protocol can yield a greater accuracy per channel invocation than the product state protocols. We show that, for some parameters and initial states, using more than two qubits and channel invocations is advantageous. These results stand in contrast to the known optimal case that uses pure initial states and a single channel invocation on a pair of entangled qubits. 
\end{abstract}

\pacs{03.65.Ta, 03.67.-a,03.65.Ud}

\maketitle


\section{Introduction}
\label{sec:intro}

Quantum parameter estimation, originally motivated by optical metrology~\cite{caves80,caves81,shapiro91}, considers how to estimate parameters associated with the evolution of quantum systems by physical means. A theory of quantum estimation, derived from the laws of quantum physics and statistics, has been developed and applied to various evolution processes yielding important results for various metrology situations~\cite{helstrom76,caves80,caves81,braunstein94,huelga97,sarovar06,giovannetti06,paris09,kok10,giovannetti11}.

We consider the single qubit depolarizing channel, which maps a qubit, initially in the state $\rhoop$, via
\begin{equation}
	\rhoop \stackrel{\hat{\Gamma}(\lambda)}{\mapsto}  \rhof(\lambda) := \frac{1-\lambda}{2} \Trace{[\rhoop]} \; \hat{I} + \lambda \rhoop 
	\label{eq:depol}
\end{equation}
where $ 0 \leqslant \lambda \leqslant 1.$ The aim of a physical estimation protocol would be to determine $\lambda$ as accurately as possible by subjecting a collection of qubits to this evolution. 

The depolarizing channel is interesting for various reasons. First, depolarization is a standard model of certain noise processes and is of general interest for quantum information processing~\cite{bennett96,adami97,nielsen00,kok10,cafaro10}. Specific examples appear in nuclear magnetic resonance (NMR)~\cite{knill01,ryan09} and optical quantum information processing~\cite{banaszek04,donascimento05,karpinski08}.

Second, quantum parameter estimation reveals fundamental and quantifiable differences with classical approaches when manifestly quantum resources such as entanglement are used. For example, in the estimation of the phase associated with a unitary parameter, classical approaches yield an uncertainty that scales as ${\cal O}(1/\sqrt{n})$ where $n$ is the number of times the channel is invoked, whereas using entangled states can result in a lower uncertainty that scales as ${\cal O}(1/n)$, thus pointing to a measurable gain provided by entanglement~\cite{giovannetti06,zwierz10,zwierz12}. 

Depolarizing channel parameter estimation has been investigated previously in the context of finding optimal estimation protocols~\cite{fujiwara01,sasaki02,fujiwara03,frey11} where any possible initial state is assumed to be available.  It emerges that the optimal estimation protocol requires qubit pairs, each initially in an entangled pure state, and that states displaying entanglement or correlations amongst more than two qubits will not give any advantages to estimation accuracy~\cite{fujiwara03}. 

We consider various estimation protocols for the constrained situation where \emph{when the available initial state of each qubit is mixed.} This situation arises in room temperature solution state NMR, where the initial state for any nuclear spin is highly mixed~\cite{nielsen00}. We ask whether, for mixed initial state situations, protocols involving correlated states can yield gains in estimation accuracy compared to protocols that use product states and whether increasing the number of correlated qubits beyond two is advantageous. While we will consider these questions for situations involving initial states with all possible ranges of mixedness, we will focus on the special case where the initial state of each qubit is highly mixed. Our approach is motivated by similar work for other types of channels where gains were found when using correlated states~\cite{modi11,collins13} and is also inspired by various other approaches to investigating parameter estimation with mixed initial states~\cite{dariano05,boixo08,braun10,datta11,pinel13}.

This article is organized as follows. Section~\ref{sec:genestimation} offers a review of quantum estimation theory. Section~\ref{sec:indest} applies this to the simplest depolarizing channel parameter channel estimation protocol, establishes a benchmark against which other schemes are to be compared and reviews existing results for parameter estimation using pure initial states. Section~\ref{sec:indchanneluseest} considers mixed initial state estimation protocols where there are never any correlations between qubits. Section~\ref{sec:correst}, which contains the key general results of our article, considers an estimation protocol that employs correlated states. Section~\ref{sec:weakpol} applies these to the special case where the available initial states are highly mixed. Finally section~\ref{sec:correlation} briefly discusses the roles of entanglement and quantum discord in the protocol that uses correlated states.



\section{Quantum parameter estimation}
\label{sec:genestimation}

\begin{figure}[h]
  \includegraphics[scale=.625]{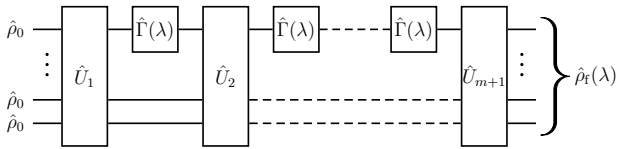} 
  \caption{Quantum estimation procedure using $n$ quantum systems, each represented by a horizontal line. This procedure uses $m$ invocations of the operation $\hat{\Gamma}(\lambda),$ and parameter-independent unitary transformations $\hat{U}_1, \ldots,  \hat{U}_{m+1}.$ The initial state of each quantum system is $\rhoo$ and the pre-measurement state of the collection of systems is $\rhof(\lambda).$ Measurements on individual systems follow $\hat{U}_{m+1}.$
           \label{fig:generalscheme}}
 \end{figure}

Physical parameter estimation requires that suitably prepared quantum systems be subjected to the operation $\hat{\Gamma}(\lambda),$ which is dependent on a real parameter $\lambda$ that is to be estimated.  The generic scheme for doing this, illustrated in Fig.~\ref{fig:generalscheme}, assumes that $m$ copies of the operation, or channel, are available. These may be invoked on one or more of $n$ systems with invocations surrounded by a series of parameter independent unitary transformations.  
Each quantum system is prepared in a known initial state $\rhoo$. After the final unitary, a measurement is chosen and performed on each system and the measurement outcomes are recorded. The final pre-measurement state $\rhof(\lambda)$ may depend on the parameter and thus so will the probabilities with which the measurement outcomes occur. Thus information about $\lambda$ might be inferred from the measurement outcomes by classical data processing, thus yielding an estimate for the parameter. 

The fact that measurement outcomes in quantum physics are usually statistically distributed implies that different runs of the same estimation procedure will yield estimates that fluctuate about a mean value. The aim of quantum parameter estimation is to tailor the process by choosing the initial state $\rhoo$, the parameter independent unitaries  $\hat{U}_1, \ldots,  \hat{U}_{m+1}$ and the final measurements so that the estimates will display minimal fluctuations about the actual parameter value.

The accuracy of the estimation is quantified via these statistical fluctuations as follows. Given a set of measurement outcomes, one on each qubit and denoted $\{ x_1, \ldots, x_n\}$, an estimate for the parameter $\tilde{\lambda}$ is attained via $\tilde{\lambda}= \lambdaest(x_1, \ldots, x_n)$ where $\lambdaest$ is an estimator function. Fluctuations in these measurement outcomes  from one run of the estimation procedure to another generate fluctuations  in $\tilde{\lambda}$. The extent to which this occurs depends on the construction of the estimator function. Key requirements for an estimation protocol will be that the estimator is unbiased, i.e. $\expv{\estimate} = \lambda$, and that it minimizes the variance, 
\begin{equation}
	\variance{(\estimate)}:= \left< \left( \estimate - \expv{\estimate} \right)^2\right>,
\end{equation}
where the angle brackets indicate the mean over all possible measurement outcomes. The variance will depend on the choice of estimator. However, a key result~\cite{cramer46,paris09,oloan10} is that \emph{the variance on any estimator  is bounded from below via the Cram\'{e}r-Rao bound}
\begin{equation}
	\variance{(\estimate)} \geqslant \frac{1}{F(\lambda)}
  \label{eq:classicalcrb}
\end{equation}
where 
\begin{equation}
	F(\lambda):= \int 
	             \left[ 
	               \frac{\partial \ln{p(x_1, \ldots x_n|\lambda)}}{\partial \lambda}
	             \right]^2\;
	             \;
	             \mathrm{d}x_1 \ldots \mathrm{d}x_n.
	\label{eq:classicalfisher}
\end{equation}
is the (classical) Fisher information associated with the probability distribution for the measurement outcomes $p(x_1, \ldots x_n|\lambda)$. This is independent of the choice of estimator. Quite generally, there is always an estimator~\cite{cramer46}, which asymptotically saturates the bound of Eq.~\eqref{eq:classicalcrb}. Thus in classical parameter estimation the Fisher information quantifies the accuracy of any possible estimate; the larger $F(\lambda)$, the better the estimate.  

In classical estimation, it is easily shown~\cite{oloan10} that if the same estimation trial is repeated $m$ times independently, the the Fisher information becomes $mF(\lambda),$ where $F(\lambda)$ is the Fisher information for a single trial. Thus
the Cram\'{e}r-Rao bound implies a reduction in the variance by a factor of $m$, yielding the typical classical reduction in uncertainty of the estimate by a factor of $\sqrt{m}.$

In quantum evolution processes the probability distribution for the measurement outcomes depends on the  type of measurement. Thus, given many possible choices of measurement, there will be many possible values for the classical Fisher information for the same evolution process. Quantum estimation aims for a measurement which results in the largest classical Fisher information. This is constrained and simplified by the \emph{quantum Cram\'{e}r-Rao bound,} which states that, for any measurement the resulting classical Fisher information satisfies
\begin{equation}
	F(\lambda) \leqslant H(\lambda)
	\label{eq:quantumcrb}
\end{equation}
where $H(\lambda)$ is the quantum Fisher information and is \emph{independent of the choice of measurement}~\cite{braunstein94,paris09,oloan10,watanabe14}. The quantum Fisher information depends only on $\rhof(\lambda)$ and is calculated via~\cite{paris09}
\begin{equation}
	H(\lambda) = \Trace{\left[ \rhof(\lambda) \score^2(\lambda)\right]}
	\label{eq:quantumfisher}
\end{equation}
where the symmetric logarithmic derivative (SLD) or score operator, $\score(\lambda)$ is defined implicitly via
\begin{equation}
	\frac{\partial \rhof(\lambda)}{\partial \lambda} = \frac{1}{2}\;
	                                                   \left[
	                                                     \score(\lambda)
	                                                     \rhof(\lambda)
	                                                     +
	                                                     \rhof(\lambda)
	                                                     \score(\lambda)
	                                                   \right].
  \label{eq:slddefintion}
\end{equation}
In general the SLD is not immediately apparent from $\rhof(\lambda)$; sometimes it can be computed using relatively simple algebra~\cite{collins13}. However, the SLD and quantum Fisher information can always be computed whenever an eigenvalue and eigenstate decomposition for $\rhof(\lambda)$ can be found. Specifically, if $ \rhof(\lambda) = \sum_j p_j \ket{\phi_j}\bra{\phi_j}$ then~\cite{paris09}
\begin{eqnarray}
  H & = &\sum_j p_j \left( \frac{1}{p_j} \frac{\partial p_j}{\partial \lambda }\right)^2  \nonumber \\
     &    & + 2 \sum_{j,k}\frac{(p_j - p_k)^2}{p_j + p_k} 
			    \left\lvert \bra{\phi_j} \frac{\partial \ket{\phi_k}}{\partial \lambda} \right\rvert^2.
	\label{eq:qfieigenval}
\end{eqnarray}

Quite generally, there exists a measurement procedure which asymptotically attains the quantum Cram\'{e}r-Rao bound~\cite{barndorff00}. Thus the quantum Fisher information quantifies the accuracy of the quantum estimation process. The main task in quantum parameter estimation is to provide a protocol giving a final pre-measurement state $\rhof(\lambda)$ with the maximum quantum Fisher information; this approach has been widely adopted elsewhere~\cite{braunstein94,fujiwara01,fujiwara03,ballester04,fujiwara04,hotta05,giovannetti06,sarovar06,ji08,paris09,oloan10,datta11,escher11,frey11,giovannetti11,modi11,zwierz12,collins13}.

Before continuing, we remark that a useful computational alternative to Eq.~\eqref{eq:quantumfisher} is~\cite{collins13}
\begin{equation}
 H(\lambda) = \Trace{\left[  \frac{\partial \rhof(\lambda)}{\partial \lambda} \score(\lambda) \right]}.
            \label{eq:quantumfishertwo}
\end{equation}

The quantum Fisher information will depend on the resources, such as the numbers of qubits,  channel invocations, single qubit unitaries or entangling unitaries that are used and various competing estimation protocols can only be compared in terms of their resource use. In this article we take a common approach, which regards the number of channel invocations $m$ in the circuit of Fig.~\ref{fig:generalscheme} as the only important resource cost~\cite{vandam07,paris09,giovannetti11,zwierz10,zwierz12}. In this context, the goal of any quantum estimation protocol is to produce a final state with the maximum quantum Fisher information per channel invocation. 

The initial state for the entire system can be assumed to be a product of states for the individual systems. If necessary, any of $\hat{U}_1, \ldots,  \hat{U}_{m+1}$ can be used to generate entangled or otherwise correlated states as the procedure unfolds. We can then consider two broad classes of estimation schemes. In the first the unitaries $\hat{U}_1, \ldots,  \hat{U}_{m+1}$ never produce any correlation between the individual systems and the entire system evolves through a series of product states. Various classical approaches to estimation, including repetition and averaging, fall into this realm. In the second, some of the unitaries $\hat{U}_1, \ldots,  \hat{U}_{m+1}$ produce states where there is some form of quantum correlation, such as entanglement, between the individual systems.  Frequently quantum estimation aims to show that using such correlated states results in a parameter estimate with a greater quantum Fisher information per channel invocation~\cite{giovannetti06,fujiwara01,fujiwara04,modi11,collins13} than is possible with non-correlated states; indirectly this indicates the utility of  quantum resources for information processing.

\subsection{Initial state considerations}

Within the framework of Fig.~\ref{fig:generalscheme}, the typical task is to find the initial state $\rhoo$ and the unitaries $\hat{U}_1, \ldots,  \hat{U}_{m+1}$ that produce a final state $\rhof(\lambda)$ that yields the maximum quantum Fisher information per channel invocation. The convexity of the quantum Fisher information guarantees that this will be attained by using a pure initial state~\cite{fujiwara01,hotta05}, and most investigations of depolarizing channel parameter estimation have considered this case~\cite{fujiwara01,sasaki02,fujiwara03,frey11}. However, it may be that case that the available initial states are not pure. Our aim is to consider various estimation protocols when the available initial states are limited to those that are mixed. 

The general initial state of a single qubit is
\begin{equation}
  \rhoo = \frac{1}{2}\; \left( \iop  + \myvector{r} \cdot \hat{\myvector{\sigma}} \right)
	\label{eq:genpureinitial}
\end{equation}
where $\myvector{r} \cdot \hat{\myvector{\sigma}} = r_x \sigmax + r_y \sigmay + r_z \sigmaz$ and $r_x, r_y, r_z$ are the components of a three dimensional real vector, $\myvector{r}$. The magnitude of this vector $r := \sqrt{r_x^2 + r_y^2 + r_z^2}$ is called the \emph{polarization or purity} of the state and it satisfies $0 \leqslant r \leqslant 1.$ When $r=1$ the state is pure and as $r$, which we use to quantify the mixedness or purity of the state, decreases the state becomes more mixed. 

Thus the principal issue of such mixed state quantum estimation is: \emph{given that the available qubits have initial states each with polarization $r$, how do the accuracies, quantified by the quantum Fisher information, of various parameter estimation protocols compare?} Note that, in this approach, the orientation of the polarization vector $\myvector{r}$ is irrelevant, as this can be modified at no cost with a single qubit unitary without changing the polarization $r$. 


\section{Depolarizing channel parameter estimation: background and baselines}
\label{sec:indest}

\begin{figure}[ht]
  \includegraphics[scale=.80]{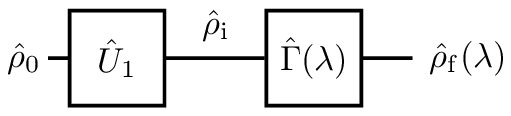} 
  \caption{ SQSC protocol. The qubit is initially in initial state $\rhoo$. A parameter independent unitary transformation $\hat{U}_1$ precedes the single channel invocation, producing a channel input state, $\rhoi$. A measurement follows the channel invocation.  
           \label{fig:sqsc}}
 \end{figure}

The simplest parameter estimation scheme, which also forms the baseline against which all other schemes may be compared, is that where the channel is invoked once on a single qubit and is described in Fig.~\ref{fig:sqsc}. This will be termed the single qubit, single channel (SQSC) protocol.

The unitary $\hat{U}_1$ of Fig.~\ref{fig:sqsc} only changes the orientation of $\myvector{r}$, leaving the polarization unaltered and thus can be omitted. Then Eq.~\eqref{eq:depol} yields that the state after one channel invocation is $\rhof = \left( \iop  + \lambda \myvector{r} \cdot \hat{\myvector{\sigma}} \right)/2$. The quantum Fisher information can be computed by first determining the SLD using Prop.~1 of~\cite{collins13} and then substituting into Eq.~\eqref{eq:quantumfishertwo}. This proposition (appropriately corrected) states that if $\alpha: = \Trace{(\rhof^2)} - (\Trace{\rhof})^2 = 0$ then 
$\score = \left[ 2 \frac{\partial \rhof}{\partial \lambda} 
                       - \frac{\partial \ln{\lvert \Trace{\rhof}\rvert} }{\partial \lambda} \rhof
               \right]/\Trace{\rhof}$
while if $\alpha \neq 0$ then
$\score = \left[ 2 \frac{\partial \rhof}{\partial \lambda} 
                       - \frac{\partial \ln{\lvert\alpha \rvert} }{\partial \lambda} \rhof
               \right]/\Trace{\rhof}
							+ \frac{\partial}{\partial \lambda} \ln{(\lvert \alpha/\Trace{\rhof} \rvert) } \iop.$
Straightforward calculation gives that $\Trace{\rhof} = 1$ and $ \alpha= (\lambda^2 r^2 - 1)/2.$ The only circumstances under which $\alpha =0$ are when $r=1$ \emph{and} $\lambda =1.$ We henceforth only consider $\lambda < 1.$ It follows that, \emph{for a single channel invocation on the state of Eq.~\eqref{eq:genpureinitial}} the SLD is
\begin{equation}
  \score = -\frac{\lambda r^2}{1- \lambda^2 r^2 } \iop + \frac{1}{1-\lambda^2 r^2}\; \myvector{r} \cdot \hat{\myvector{\sigma}}.
\end{equation}
Substituting into Eq.~\eqref{eq:quantumfishertwo} then gives that \emph{the quantum Fisher information for the SQSC protocol using an initial state with polarization $r$} is
\begin{equation}
  \Hs = \frac{r^2}{1 - \lambda^2 r^2}.
	\label{eq:qfisingle}
\end{equation}
This is the baseline quantum Fisher information against which all other estimation schemes are to be compared. 

\subsection{Pure Initial State Estimation Schemes}

We briefly digress to discuss the previously investigated~\cite{fujiwara01,sasaki02,fujiwara03,frey11} extreme case, where the initial states are pure.  If it were possible to modify the initial state polarization, then Eq.~\eqref{eq:qfisingle}, shows that, in accordance with general results~\cite{fujiwara01,hotta05}, the largest quantum Fisher information would be attained for a pure input state, i.e.\ when $r=1,$ giving
\begin{equation}
  \Hspure= \frac{1}{1 - \lambda^2}.
	\label{eq:qfisinglepure}
\end{equation}

Usually the key issue of interest in quantum parameter estimation is whether a greater quantum Fisher information can be attained when entangled multiple qubits are available as input for the channel. For the depolarizing channel it has been shown that entanglement can yield a greater quantum Fisher information. Specifically if the channel is applied to one of two qubits then the optimal quantum Fisher information is  attained~\cite{fujiwara01,frey11} when the two qubits are prepared in a maximally entangled pure state, such as $\ket{\Psi_i} = \left( \ket{00} + \ket{11}\right)/\sqrt{2}$. The resulting quantum Fisher information is 
\begin{equation}
	\Hoptpure = \frac{3}{(3+\lambda)(1-\lambda)}.
	\label{eq:qfientangledpure}
\end{equation}
It is straightforward to show that $\Hoptpure > \Hspure$ whenever $\lambda>0.$ 

The remaining question is whether any additional qubits and channel invocations can yield a greater quantum Fisher information per channel invocation. Remarkably the result is~\cite{fujiwara03} that the optimal depolarizing channel parameter estimation scheme uses pairs of qubits, each in a maximally entangled pure input state. The channel is applied once to only one of each entangled pair and the resulting quantum Fisher information per channel use is that of Eq.~\eqref{eq:qfientangledpure}. Thus \emph{when qubits are available in pure initial states then entangling more than two qubits cannot yield any advantages for depolarizing channel parameter estimation.} However, we stress that this does not necessarily pertain to the situation that we address, in which the available input states are not pure.


\section{Multiple Channel Uses: Uncorrelated State Protocols}
\label{sec:indchanneluseest}

The previous discussion motivates the central questions of our work: \emph{given a collection of qubits, each initially in a state with the same polarization $r$, is it possible to enhance the parameter estimation accuracy by using multiple channel invocations or entanglement or correlations amongst more than two qubits? If so to what extent can the accuracy, as measured by the quantum Fisher information per channel use, be enhanced?}

\begin{figure}[ht]
  \includegraphics[scale=.80]{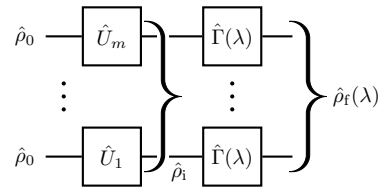} 
  \caption{Independent channel use protocol. There are $m$ single qubits, each initially in the state $\rhoo$. Qubit $j$ is subjected  to a single qubit unitary, $\hat{U}_j $, which is independent of the channel parameter, producing the input state $\rhoprep.$ Thereafter a depolarizing channel is applied once to each qubit.
           \label{fig:indchanneluse}}
 \end{figure}

A first strategy for enhancing estimation accuracy is one where the same process is repeated independently, as is done for classical repeat and average schemes. Specifically we consider the protocol of Fig.~\ref{fig:indchanneluse}, called the \emph{independent channel use protocol.} Here there are $m$ qubits, each in the initial state $\rhoo$ and subject to a parameter independent single qubit preparatory unitary. The channel is subsequently invoked exactly once on each qubit. In this protocol there are never any correlations between the qubits during the entire process; in this sense it resembles a classical approach.  
%

The quantum Fisher information is additive in such cases~\cite{oloan10,collins13} and thus the quantum Fisher information for the independent channel use protocol with $m$ channel invocations is
\begin{equation}
	H(\lambda) = m \Hs = \frac{mr^2}{1 - \lambda^2 r^2}.
\end{equation}
Although this is an improvement over the SQSC protocol, clearly the quantum Fisher information \emph{per channel use} in the independent channel use protocol is identical to that of the SQSC protocol. Thus \emph{the independent channel use protocol attains the same estimation accuracy, measured in terms of quantum Fisher information per channel invocation, as the SQSC protocol.}

\begin{figure}[ht]
  \includegraphics[scale=.80]{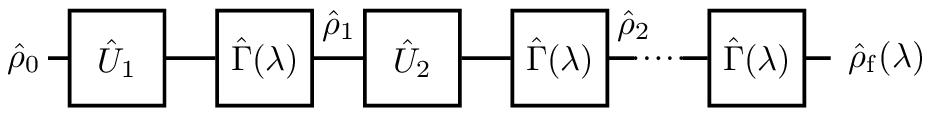} 
  \caption{Sequential channel use protocol. A single qubit, initially in the state $\rhoo$ is subjected to $m$ channel invocations. Each channel invocation is preceded a single qubit unitary, $\hat{U}_1 \ldots \hat{U_m}$, all of which are independent of the channel parameter. 
           \label{fig:seqchanneluse}}
 \end{figure}

\begin{figure}[ht]
  \includegraphics[scale=.80]{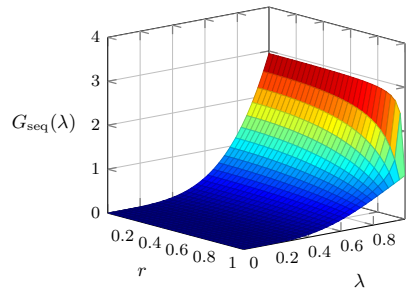} 
  \caption{Sequential channel use gain for $m=3$.
           \label{fig:gainseqplot}}
 \end{figure}

A second strategy, illustrated in Fig.~\ref{fig:seqchanneluse},  is that where \emph{only one qubit is available and the channel is invoked sequentially a total of $m$ times.} Successive channel invocations are interspersed with parameter-independent unitaries.
Assuming the standard initial state, $\rhoo$, the effect of $\hat{U}_1$ is to rotate the polarization vector, $\myvector{r} \rightarrow \myvector{r}_1$ without altering its magnitude. Then Eq.~\eqref{eq:depol} implies that the state after the first channel invocation is $\rhoone= ( \iop  +  \lambda \myvector{r}_1 \cdot \hat{\myvector{\sigma}}) /2$. Using the same argument, the state after the second channel invocation is $\rhotwo= ( \iop  +  \lambda^2 \myvector{r}_2 \cdot \hat{\myvector{\sigma}}) /2$ where $\hat{U}_2$ rotates $\myvector{r}_1 \rightarrow \myvector{r}_2.$ Continuing, the state prior to measurement is $\rhof= ( \iop  +  \lambda^m \myvector{r}_\mathrm{f} \cdot \hat{\myvector{\sigma}}) /2$ where $r_\mathrm{f}= r.$ A calculation similar to that giving Eq.~\eqref{eq:qfisingle} yields that the quantum Fisher information for $m$ sequential channel invocations on a single qubit is
\begin{equation}
  \Hseq = m^2 \frac{\lambda^{2m-2}r^2}{1 - \lambda^{2m} r^2}.
	\label{eq:qfiseq}
\end{equation}
Thus the quantum Fisher information per channel invocation is 
\begin{equation}
  \Hseqper = m \frac{\lambda^{2m-2}r^2}{1 - \lambda^{2m} r^2}.
	\label{eq:qfiseqper}
\end{equation}
To compare this to the SQSC protocol we define the \emph{sequential channel use gain} as
\begin{equation}
  \Gseq := \frac{\Hseqper}{\Hs} = m \frac{\lambda^{2m-2} - \lambda^{2m} r^2}{1 - \lambda^{2m} r^2}.
	\label{eq:gainseq}
\end{equation}
Figure~\ref{fig:gainseqplot} illustrates the typical gain, plotted here for the  $m=3$ case. This shows that the sequential channel use protocol is advantageous for certain ranges of parameter values and polarizations but disadvantageous for others. 

The sequential channel use gain decreases monotonically with $r$ and increases monotonically with $\lambda$ (these are proved in Appendix~\ref{app:gainseq}). A more detailed analysis of this separates the pure initial state case, i.e.\ $r=1$, from mixed initial state cases, i.e.\ $r<1$. For the \emph{pure initial state case,} it is shown in Appendix~\ref{app:gainseq} that 
\begin{equation}
	 0 \leqslant \Gseq \leqslant 1.
\end{equation}
 For the \emph{mixed initial state case,} it is also shown Appendix~\ref{app:gainseq} that 
\begin{equation}
	 0 \leqslant \Gseq \leqslant m.
\end{equation}
In both cases the upper bound is attained when $\lambda=1.$

It follows that \emph{if the available initial state is pure there is no advantage to the sequential channel use protocol. If the available initial state is mixed then, for certain initial polarizations and the parameter values, there can be an advantage to the sequential channel use protocol.}

It remains to assess whether for mixed initial states, there is any advantage to increasing the number of channel invocations. To do so, we compare the quantum Fisher information per channel use of Eq.~\eqref{eq:qfiseqper} for $m+1$ channel invocations versus that for $m$ channel invocations. An advantage is attained when the ratio of that for $m+1$ to that for $m$ exceeds $1$, and straightforward algebra shows that this is equivalent to $r^2 \leqslant [\lambda^2(m+1) - m]/\lambda^{2m+2}$. The right hand side of this increases monotonically  with $\lambda$ and attains a maximum of $1$ when $\lambda=1$. Also it is negative when $\lambda < \sqrt{m/(m+1)}$, implying that when  $\lambda < \sqrt{m/(m+1)}$ there is no polarization for which an additional channel use will give an advantage.  Thus for any polarization there will be some parameter values for which additional invocations in the sequential channel use protocol will be advantageous. However, the range of possible parameter values for which this occurs, bounded from below by at least $\sqrt{m/(m+1)}$, shrinks with increasing channel invocations. 


\section{Multiple Qubit or Channel Uses: Correlated State Protocol}
\label{sec:correst}

An additional resource available to multiple qubit systems, is the existence of states in which there are correlations between the individual qubits and we now consider whether these can enhance the accuracy of parameter estimation for the depolarizing channel. 

\begin{figure}[ht]
  \includegraphics[scale=.80]{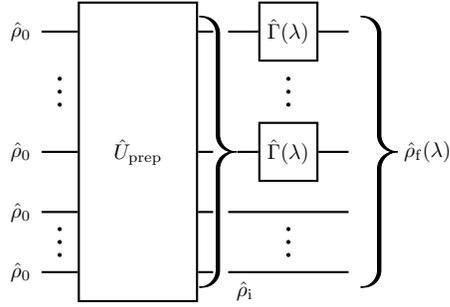} 
  \caption{Correlated state protocol considered in this article. There are $n$ single qubits, each initially in the state $\rhoo$. These are subjected  to a joint preparatory unitary, $\uprep$, producing the channel input state $\rhoprep.$ Thereafter a depolarizing channel is applied once to each of the uppermost $m$ qubits. The preparatory unitary is illustrated in Fig.~\ref{fig:uprep}.
           \label{fig:preparedscheme}}
 \end{figure}

\begin{figure}[ht]
  \includegraphics[scale=.80]{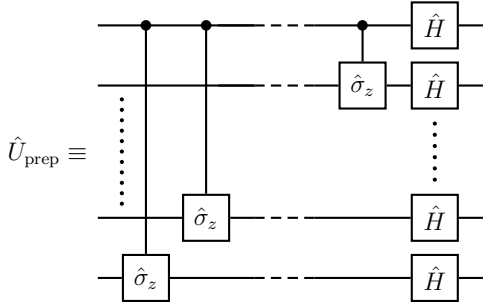} 
  \caption{The preparatory gate, $\uprep,$ consists of a collection of controlled-$Z$ operations applied to each distinct pair of qubits followed by a Hadamard transformation applied to each qubit. 
           \label{fig:uprep}}
 \end{figure}

We consider the particular correlated state protocol procedure of Fig.~\ref{fig:preparedscheme}; this was previously investigated for other channels~\cite{modi11,collins13}.  We assume that each qubit is initially in the particular state $\rhoo = (\hat{I}+ r \sigmay)/2$ with polarization $r$ and that a preparatory unitary $\uprep$, illustrated in Fig.~\ref{fig:uprep}, is subsequently applied jointly to the qubits.
Denote the density operator for the system after the preparatory gate by $\rhoprep : =  \uprep \rhoi \uprep^\dagger.$
It can be shown~\cite{collins13} that 
\begin{equation}
  \rhoprep = \sum_{x=0}^{(N-1)/2} \rhox
\end{equation}
where $N= 2^n-1$ and 
%
\begin{equation}
	\begin{split}
	  \rhox = &
		          \diag{j(x)}\;
							\biggl[
		           \ket{x}\bra{x} + 
							 \ket{N-x}\bra{N-x}
              \biggr]
		        \\
	          & 
						+ i \cdiag{j(x)}\;
							\biggl[
		           \ket{x}\bra{N-x} -
							 \ket{N-x} \bra{x}
              \biggr].
	\end{split}
 \label{eq:rhoprepped}
\end{equation}
The counter-diagonal terms of the resulting matrix of Eq.~\eqref{eq:rhoprepped} are constructed by first computing the number of zeroes $j(x)$ in the binary representation $x_n\ldots x_1$ for $x$ and then using 
 \begin{eqnarray}
         \cdiag{j} & = & \frac{1}{2^{n+1}}\; \left[ (1+r)^j(1-r)^{n-j} \right. \nonumber \\
	       &    & \left. \phantom{\frac{1}{2^{n+1}}\; [} - (1+r)^{n-j}(1-r)^j \right]. 
				 \label{eq:prepcounterdiag}
 \end{eqnarray}
The diagonal terms are
 \begin{eqnarray}
         \diag{j}& = & \frac{1}{2^{n+1}}\; \left[ (1+r)^j(1-r)^{n-j} \right. \nonumber \\
	       &    & \left. \phantom{ \frac{1}{2^{n+1}}\; [}  + (1+r)^{n-j}(1-r)^j \right]. 
				 \label{eq:prepdiag}
 \end{eqnarray}
As described previously~\cite{collins13}, this scheme results in a density operator which has supports consisting of mutually orthogonal two dimensional subspaces; in terms of matrix representations it can be expressed as a sum of $2\times 2$ matrices acting on distinct vector spaces. 

When the initial states are pure, i.e.\ $r=1,$ the diagonal and counter-diagonal terms are all zero except $c_0 = -c_n = -1/2$ and $d_0 = d_n = 1/2$ and $\rhoprep$ is the density operator corresponding to the GHZ-type state $(\ket{0\ldots 0} + \ket{1\ldots 1})/\sqrt{2}.$

Suppose that the channel acts \emph{once on each of $m$ qubits} corresponding to the least significant (or rightmost) digits in the computational basis representation. The action of the channel on multiple qubits is readily described by noting that Eq.~\eqref{eq:depol} implies that, for a single qubit, 
\begin{subequations}
 \label{eq:depoloncompbasis}
 \begin{eqnarray}
         \ket{x}\bra{y} & \stackrel{\hat{\Gamma}(\lambda)}{\mapsto} &  \lambda  \ket{x}\bra{y}  \quad \textrm{and} \label{eq:depoloncompbasiscdiag} \\
	       \ket{x}\bra{x} & \stackrel{\hat{\Gamma}(\lambda)}{\mapsto} &  p \ket{x}\bra{x} + q \ket{\overline{x}}\bra{\overline{x}} \label{eq:depoloncompbasisdiag} 
 \end{eqnarray}
\end{subequations}
where $\ket{x}$ and $\ket{y}$ are computational basis states such that $x \neq y$ and $\overline{x} := 1-x.$ Here $p:=(1+\lambda)/2$ and  $q:=(1-\lambda)/2$. A general state of multiple qubits can be expressed in terms of linear combinations and tensor products  of basic operators such as $\ket{x}\bra{y}$ and thus Eqs.~\eqref{eq:depoloncompbasis} can be used to describe the action of any number of channels acting on multiple qubits.

Eq~\eqref{eq:depoloncompbasiscdiag} implies the counter-diagonal terms of the matrix representing $\rhof(\lambda)$ are attained from those of $\rhoprep$ by multiplication by $\lambda^m.$ Eq.~\eqref{eq:depoloncompbasisdiag} implies that the diagonal terms of the matrix representing $\rhof(\lambda)$  result from rearrangements of those of $\rhoprep$ multiplied by various factors of $p$ and $q$. The remaining terms are again all $0$. An additional simplification, demonstrated in appendix~\ref{app:rhofdiag} is that $\bra{N-x} \rhof(\lambda)  \ket{N-x} = \bra{x} \rhof(\lambda)  \ket{x}.$ Thus the final density operator has a form similar to that for $\rhoprep$, i.e.\
\begin{equation}
  \rhof(\lambda) = \sum_{x=0}^{(N-1)/2} \rhofx
	\label{eq:rhofinalsum}
\end{equation}
where
\begin{equation}
	\begin{split}
	  \rhofx = &
		          \diagf{x}\;
							\biggl[
		           \ket{x}\bra{x} + 
							 \ket{N-x}\bra{N-x}
              \biggr]
		        \\
	          & 
						+ i \cdiagf{x}\;
							\biggl[
		           \ket{x}\bra{N-x} -
							 \ket{N-x} \bra{x}
              \biggr].
	\end{split}
 \label{eq:rhofinal}
\end{equation}
Eq.~\eqref{eq:depoloncompbasiscdiag} yields
\begin{equation}
	\cdiagf{x} = \lambda^m \cdiag{j(x)}
	\label{eq:cdiagfinal}
\end{equation}
for the counter-diagonal terms where $j(x)$ is the number of zeroes in the bit string for $x$. 

Determining the diagonal terms is further simplified by noting that the preparatory unitary is invariant under interchange of any two qubits and that this is also true of the channel operation provided that the channel either acts on each qubit that is interchanged or that it acts on neither of the qubits that are interchanged. The invariance of the preparatory unitary is readily apparent from its structure, as shown in Fig.~\ref{fig:uprep}. More precisely, suppose that $\hat{S}_{j^\prime j}$ is the unitary operation which swaps the qubits labeled $j$ and $j^\prime$. Then if $1\leqslant j<j^\prime \leqslant m$  (recall that the channel acts on the rightmost $m$ bits), 
\begin{equation}
	\bra{y} \hat{S}_{j^\prime j}^\dagger \rhof \hat{S}_{j^\prime j} \ket{x} = \bra{y} \rhof \ket{x} 
\end{equation}
with an identical result when $m+1\leqslant j<j^\prime \leqslant n$. Thus if the bit string for $x$ is $x_n\ldots x_{m+1}x_m \ldots x_1$, then $\diagf{x} =  \bra{x} \rhof \ket{x} $ is invariant under the interchange of any two of $\{x_n, \ldots, x_{m+1}\}$ and is also invariant under the interchange  any two of $\{x_m, \ldots, x_1\}$ (it is not in general invariant under interchanges between one element from one of these sets and another from the other set). For example, if the channel is applied the two least significant qubits out of a total of five qubits then $\diagf{6} = \diagf{5}$ (or, using a binary representation for the index subscript, $\diagf{00110} = \diagf{00101}$) and $\diagf{26} = \diagf{21}$ (or $\diag{\textrm{f}\; 11010} = \diag{\textrm{f}\; 10101}$). A consequence of this is that $\diagf{x}$ only depends on the number of zeroes in the leftmost $n-m$ bits of the binary representation of $x$ and also the number of zeroes in the rightmost $m$ bits of the binary representation of $x$. This consideration and a derivation, provided in appendix~\ref{app:rhofdiagterms}, yields
\begin{equation}
	\diagf{x} = \sum_{k=0}^{m} q^k p^{m-k} 
	                                  \sum^{l_\mathrm{max}}_{l=l_\mathrm{min}}\;
																		\binom{v}{l}
																		\binom{m-v}{k-l}\;
																		\diag{u+v+k-2l}
	\label{eq:diagfinal}
\end{equation}
 where $u = u(x)$ is the number of zeroes  in the leftmost $n-m$ bits (i.e.\ $x_n\ldots x_{m+1}$) of the binary representation of $x$, $v=v(x)$ is the number of zeroes in the rightmost $m$ bits (i.e.\ $x_m \ldots x_1$) of the  binary representation of $x$,  $l_\mathrm{min} =\max{(k+v-m,0)},$ and $l_\mathrm{max} =\min{(k,v)}$. Clearly $\diagf{x}$ depends on $u,v$ and $m$ and $\lambda$, via $p= (1 +\lambda)/2$ and $q= (1 - \lambda)/2$.  Since $\lambda$ is implicit throughout this work and $m$ is fixed by the number of channel invocations we will find it convenient to express $\diagf{x}$ as $\diagf{x} = \diagf{}(u,v).$

According to Eqs.~\eqref{eq:rhofinalsum} and~\eqref{eq:rhofinal}, $\rhof$ is a sum of operators, each acting on a two dimensional space, which have mutually orthogonal supports. Thus Prop.~2 of~\cite{collins13} implies that the quantum Fisher information is
\begin{equation}
 \Hcorr = \sum_{x=0}^{(N-1)/2} H_x(\lambda)
  \label{eq:qfisum}
\end{equation}
where $H_x(\lambda)$ is the quantum Fisher information associated with $\rhofx.$ A derivation (see appendix~\ref{app:qfifirst}) based on this and  Eq.~\eqref{eq:qfieigenval} yields
%
\begin{eqnarray}
 \Hcorr & = & \sum_{x=0}^{(N-1)/2} \frac{2}{\diagf{x}^2 - \lambda^{2m} \cdiag{j(x)}^2 }  \nonumber \\
          &   & \! \! \!
					   \times
		          \left[ \diagf{x} \left(  \dotdiagf{x}^{\phantom{f}2} + m^2 \lambda^{2m-2} \cdiag{j(x)}^2 \right) \right.\nonumber \\
          &   & \; \; \; 
					      \left. 
		                  - 2 m\lambda^{2m-1} \dotdiagf{x} \cdiag{j(x)}^2
		          \right]
\label{eq:qfidepolone}
\end{eqnarray}
%
where the dot indicates differentiation with respect to $\lambda.$ Although the sum in Eq.~\eqref{eq:qfidepolone} is over $x$, the terms indexed as such depend on $u$ and $v$ and it follows that the quantum Fisher information can be expressed as a sum over these. In appendix~\ref{app:qfifirst}  it is shown that \emph{if $m<n$ then}
%
\begin{eqnarray}
 \Hcorr & = & \sum_{v=0}^{m} \sum_{u=1}^{n-m} \binom{n-m-1}{u-1} \binom{m}{v}\;
              \frac{2}{\diagf{}^2 - \lambda^{2m} \cdiag{u+v}^2 } \nonumber \\
     &   &  \! \! \!
		           \times \left[ \diagf{} \left(  \dotdiagf{}^{2} + m^2 \lambda^{2m-2} \cdiag{u+v}^2 \right)  \right.  \nonumber \\
     &   & \; \; \; 
		          \left.
		                  - 2 m\lambda^{2m-1} \dotdiagf{} \cdiag{u+v}^2
		          \right]
\label{eq:qfidepoltwomlessthann}
\end{eqnarray}
%
where $\diagf{} = \diagf{}(u,v)$ is given by Eq.~\eqref{eq:diagfinal}.  \emph{If $n=m$} then 
%
\begin{eqnarray}
 \Hcorr& = & \sum_{v=1}^{n} \binom{n-1}{v-1}\;
              \frac{2}{\diagf{}^2 - \lambda^{2m} \cdiag{v}^2 }  \nonumber \\
     &   & \! \! \!
		          \times \left[ \diagf{} \left(  \dotdiagf{}^{2} + m^2 \lambda^{2m-2} \cdiag{v}^2 \right) \right. \nonumber \\
     &   & \; \; \; 
		          \left.
		                  - 2 m\lambda^{2m-1} \dotdiagf{} \cdiag{v}^2
		          \right]
\label{eq:qfidepoltwomequalsn}
\end{eqnarray}
%
where $\diagf{} = \diagf{}(0,v)$ is again given by Eq.~\eqref{eq:diagfinal}. In both cases $\cdiag{j}$ is given by Eq.~\eqref{eq:prepcounterdiag}. These represent the simplest expressions that we have been able to obtain for the quantum Fisher information for the correlated state protocol. 

The quantum Fisher information per channel invocation is
\begin{equation}
  \Hcorrper = \frac{\Hcorr}{m}.
	\label{eq:qficorrper}
\end{equation}
We aim to compare the performance of the correlated state protocol to that of the SQSC protocol \emph{in terms of the quantum Fisher information per channel use and given that the initial qubit polarizations are the same.} We define the \emph{correlated state protocol quantum Fisher information gain} as
\begin{equation}
	\Gcorr := \frac{\Hcorrper}{\Hs}
	\label{eq:gaindef}
\end{equation}
and whenever this exceeds $1$, the correlated state protocol outperforms the SQSC protocol.
 
The expression for the quantum Fisher information is a rational polynomial in $r$ and $\lambda$ and even with as few as two qubits, this is too complicated to yield insight into the estimation accuracy for the correlated state protocol for all values of polarization and the parameter. While the same applies to the gain, aspects of its behavior can be illustrated graphically. 

\begin{figure}[ht]
  \includegraphics[scale=.80]{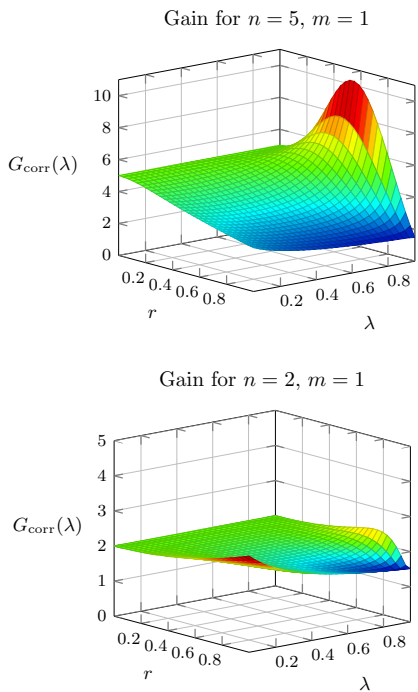} 
  \caption{Gain using one channel invocation. Both plots are over the range $0.05 \leqslant \lambda \leqslant 0.95$ and $0 \leqslant r \leqslant 1.$
           \label{fig:gainplotssingle}}
 \end{figure}

First, suppose that there is only one channel invocation ($m=1$). Fig.~\ref{fig:gainplotssingle} illustrates the gain for the examples $n=2$ and $n=5$. 
These suggest that when the channel is invoked once only, the correlated state protocol will be more accurate than the SQSC protocol, assuming the same initial state polarization for both. However, a general proof of this is elusive; the methods of~\cite{collins13} used to prove a similar fact for the phase-flip channel are not immediately applicable here. 

\begin{figure}[ht]
  \includegraphics[scale=.80]{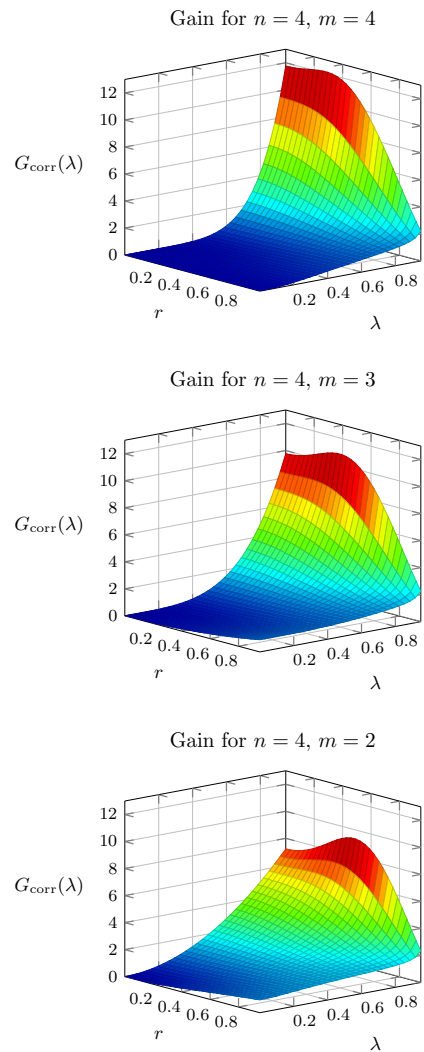} 
  \caption{Gain using multiple channel channel invocations on $4$ qubits. All plots are over the range $0\leqslant \lambda, r \leqslant 0.95$.
           \label{fig:gainplotsmultiple}}
 \end{figure}

Second, suppose that there are multiple channel invocations ($m>1$). Fig.~\ref{fig:gainplotsmultiple} illustrates the correlated state gain of Eq.~\eqref{eq:gaindef} for various numbers of channel invocations on $4$ qubits. 
These show that when the channel is invoked multiple times, there are polarizations and parameter values for which the the correlated state protocol will be more accurate than the SQSC protocol but there are also parameter values for which it is inferior. 

Given that the channel is invoked multiple times, the correlated state protocol can also be compared to the sequential channel protocol that uses the same number of channel invocations. We define the correlated vs sequential channel gain as
\begin{equation}
	\Gcorrseq := \frac{\Hcorrper}{\Hseqper}
	\label{eq:gaincorrseqdef}
\end{equation}
where the number of channel invocations is the same for the two protocols. 
 \begin{figure}[ht]
  \includegraphics[scale=.50]{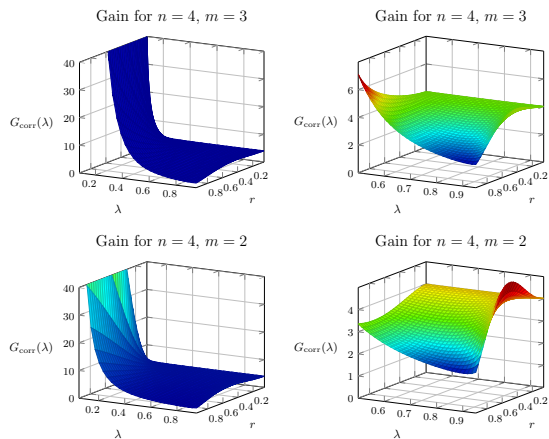} 
  \caption{Correlated vs sequential channel gain using multiple channel channel invocations on $4$ qubits. Plots on the left are over the range $0.05\leqslant \lambda, r \leqslant 0.95$. Plots on the right provide details for $\lambda \geqslant 0.5.$ 
           \label{fig:gaincorrmultvsseq}}
 \end{figure}
Figure~\ref{fig:gaincorrmultvsseq} suggests that when the channel is invoked a fixed number of times, the correlated state protocol is more accurate than the sequential protocol, using the same number of channel invocations, over all parameter values and for all polarizations. 


\section{Weak initial polarization cases}
\label{sec:weakpol}

In some situations the initial state polarization is very small, i.e.\ $r \ll 1.$ This is true of room-temperature solution state NMR~\cite{nielsen00}, where typically $r \approx 10^{-4}.$ 

When \ $r \ll 1$ the SQSC protocol quantum Fisher information becomes
\begin{equation}
  \Hs \approx r^2
	\label{eq:qfisingleapprox}
\end{equation}
and the sequential channel use protocol quantum Fisher information per channel use becomes
\begin{equation}
  \Hseqper \approx m \lambda^{2m-2}r^2,
	\label{eq:qfiseqperapprox}
\end{equation}
each to lowest order in $r.$
In appendix~\ref{app:smallrqfi} we show that the correlated state quantum Fisher information per channel use becomes
\begin{equation}
  \Hcorrper \approx mn \lambda^{2m-2}r^2,
	\label{eq:qficorrperapprox}
\end{equation}
to lowest order in $r$. 

\begin{figure}[ht]
  \includegraphics[scale=.80]{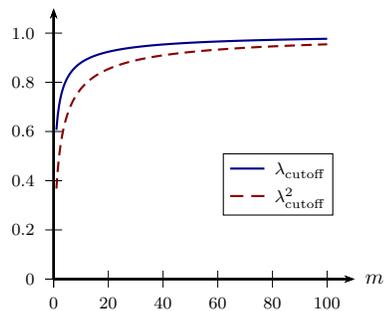} 
  \caption{Parameter cutoff function and its square versus $m$.  
           \label{fig:lambdacutoff}}
 \end{figure}

Here the sequential channel use gain approximates to
\begin{equation}
  \Gseq \approx m \lambda^{2m-2}.
	\label{eq:gainseqperapprox}
\end{equation}
Eq.~\eqref{eq:gainseqperapprox} shows that the gain exceeds $1$ when the depolarizing channel parameter satisfies  $\lambda \geqslant \lambdacutoffseq (m)$ where $\lambdacutoffseq (m) := m^{1/(2-2m)}$ is the characteristic parameter cutoff. The characteristic parameter cutoff, plotted in Fig.~\ref{fig:lambdacutoff}, is a monotonically increasing function of $m$, with a minimum value of $e^{-1/2} =0.61,$ attained as $m \rightarrow 1$ and a maximum value of $1$ attained as $m \rightarrow \infty$. This indicates that for any given parameter value, there will be a maximum number of channel invocations beneath which the sequential channel use protocol offers enhanced parameter estimation accuracy and beyond which it will not do so. Figure~\ref{fig:lambdacutoff} shows that as $\lambda$ increases, accuracy gains will persist with increasing number of channel invocations. Note that the larger $\lambda$ is, the weaker the depolarization is, and for such weak depolarization cases enhanced estimation accuracies can be attained with multiple sequential  channel invocations.
 Thus, \emph{for low polarizations, the sequential channel use protocol can yield a greater estimation accuracy per channel invocation than the SQSC protocol.} However, this requires that the parameter value exceed the characteristic parameter cutoff and Fig.~\ref{fig:lambdacutoff} indicates that such advantages will be absent for low enough parameter values and, even when there is an advantage, increasing the number of channel invocations will eventually eliminate this advantage. 

The correlated state protocol gain (versus the SQSC protocol) approximates to 
\begin{equation}
  \Gcorr \approx m n \lambda^{2m-2}.
	\label{eq:gaincorrperapprox}
\end{equation}
We consider two cases separately.

First, with \emph{only one channel invocation the correlated state protocol offers an $n$-fold gain in accuracy regardless of the parameter value, for low initial polarization states.}  This is strikingly different from the pure state case where the optimal accuracy is attained~\cite{fujiwara03} with a single channel invocation on one of only two qubits ($n=2$ and $m=1$); by contrast when the initial state polarization is low and the channel is only invoked on one qubit, then adding more qubits always increases the estimation accuracy. 

Second, with multiple channel invocations, Eq.~\eqref{eq:gaincorrperapprox} indicates that enhanced estimation accuracy will occur if $\lambda \geqslant (mn)^{1/(2-2m)} = \lambdacutoffseq(m)\, n^{1/{(2-2m)}}.$  The right hand side forms a new cutoff which is again an increasing function of the number of channel invocations but a decreasing function of the total number of qubits. Thus increasing the number of spectator qubits while leaving the number of channel invocations fixed extends the range of parameter values over which the correlated state protocol is advantageous. Furthermore Eq.~\eqref{eq:gaincorrperapprox}  shows that adding more spectator qubits while keeping the number of channel invocations fixed will increase the quantum Fisher information per channel use; the increase can be arbitrarily large by adding sufficiently many spectator qubits. \emph{Unlike the pure initial state case, for low polarization initial states, there will always be an advantage to adding additional spectator qubits.} 

Separately, when $m<n$ we can assess whether one additional channel invocation yields any advantage. Eq.~\eqref{eq:qficorrperapprox} implies that starting with $m$ channel invocations and adding one more invocation increases the quantum Fisher information if $m < \lambda^2/(1-\lambda^2)$, decreases it if  $m > \lambda^2/(1-\lambda^2)$ and makes no difference if $m = \lambda^2/(1-\lambda^2).$ Thus the optimal number of channel invocations in the correlated state protocol is $\mopt = \lfloor  \lambda^2/(1-\lambda^2) \rfloor +1,$ unless $\lambda^2/(1-\lambda^2)$ is an integer, in which case,  $\mopt =  \lambda^2/(1-\lambda^2).$ Examples of this and the associated gains are provided in Table~\ref{tab:optinvocations}.
\begin{table}[h]
 \begin{ruledtabular}
 \begin{tabular}{ccd}
   $\lambda$ & $\mopt$ & \multicolumn{1}{c}{\text{Optimal gain}}\\
	 \hline
	 $0.700$ & $1$& 1.00\, n\\
	 $0.800$ & $2$& 1.28\, n\\
	 $0.900$ & $5$& 2.15\, n\\
	 $0.950$ & $10$& 3.97\, n\\
	 $0.990$ & $50$& 18.67\, n \\
	 $0.995$ & $100$& 37.07\, n
 \end{tabular}
 \end{ruledtabular}
 \caption{Optimal number of correlated state protocol channel invocations and optimal gain, i.e.\ Eq.~\eqref{eq:gaincorrperapprox} evaluated at $\mopt$, whenever $m<n.$
          \label{tab:optinvocations}
					}
\end{table}
 When $\lambda$ is close enough to $1$, this offers considerable opportunity for improved accuracy. Again  \emph{unlike the pure initial state case, for low initial polarization, increasing the number of channel invocations in the correlated state protocol can result in enhanced estimation accuracy, per channel use.} 

In many physical situations every qubit is subjected to the depolarizing channel. When the channel is invoked on all qubits of the correlated state protocol ($m=n$), the gain becomes $\Gcorr \approx m^2 \lambda^{2m-2}$ and this implies that the correlated state protocol provides enhanced accuracy whenever $\lambda \geqslant \lambdacutoffseq^2(m).$ This lower bound is plotted in Fig.~\ref{fig:lambdacutoff}. This is again an increasing function of $m$, approaching $e^{-1} = 0.37$ as $n=m\rightarrow 1$. The slightly lower cutoff implies that there is somewhat greater scope for enhanced parameter estimation when all qubits are subjected to the channel. However, for a given parameter value, this cutoff establishes an upper bound on the number of qubits/channel invocations for which the correlated state protocol provides any gain. 

We can again ask when adding one more qubit and channel invocation gives any advantage. Equation~\eqref{eq:gaincorrperapprox} with $m=n$ implies that if there are originally $m$ channel invocations then adding one more gives a greater quantum Fisher information per channel use if $m \leqslant \mopt $ where $\mopt= \lfloor  \lambda/(1-\lambda) \rfloor+1$ unless $\lambda/(1-\lambda)$ is an integer, in which case, $\mopt= \lambda/(1-\lambda).$ Examples of this and the associated optimal gains are provided in Table~\ref{tab:optinvocationsall}.
\begin{table}[h]
 \begin{ruledtabular}
 \begin{tabular}{ccd}
   $\lambda$ & $\mopt$ & \multicolumn{1}{c}{\text{Optimal gain}}\\
	 \hline
	 $0.500$ & $1$& 1.00 \\
	 $0.700$ & $3$& 2.16 \\
	 $0.900$ & $9$& 15.01\\
	 $0.950$ & $19$& 56.96 \\
	 $0.970$ & $33$& 155.03\\
	 $0.990$ & $100$& 1367.00
 \end{tabular}
 \end{ruledtabular}
 \caption{Optimal number of correlated state protocol channel invocations and optimal gain, i.e.\ Eq.~\eqref{eq:gaincorrperapprox} evaluated at $\mopt$, whenever the channel is invoked once on each qubit ($m=n$). 
          \label{tab:optinvocationsall}
					}
\end{table}
Clearly, for relatively weak depolarizing channels, i.e.\ $\lambda \approx 1,$ substantial enhancements of estimation accuracy are possible with the correlated state protocol.

Finally, the correlated state protocol and sequential channel protocol can be compared in the low initial polarization limit, given the same number of channel invocations. Here the correlated vs sequential channel gain as
\begin{equation}
	\Gcorrseq \approx n
	\label{eq:gaincorrseqapprox}
\end{equation}
and thus when $r \ll 1,$ the correlated state protocol will always offer a better estimation accuracy than the sequential channel protocol provided that the number of channel invocation for each is the same.


\section{Quantum Correlations in the Correlated State Protocol}
\label{sec:correlation}

The presence of quantum correlations such as entanglement or discord in the correlated state protocol may be assessed for the $n=2$ along the same lines as was done for the Pauli channel parameter estimation of~\cite{collins13}. Then Eqs.~\eqref{eq:rhofinal}, \eqref{eq:cdiagfinal} and \eqref{eq:diagfinal} yield that with $m$ channel invocations,
\begin{equation}
	\rhof = \frac{1}{4}
	        \begin{pmatrix}
	         1+\lambda^m r^2 & 0 & 0 & 2 i r\lambda^m\\
					 0 & 1-\lambda^m  r^2 & 0 & 0\\
					 0 & 0 &1-\lambda^m  r^2  & 0 \\
					 -2 i r\lambda^m  & 0 & 0 & 1+\lambda^m r^2
	        \end{pmatrix},
	\label{eq:finalrhofortwo}
\end{equation}
in the computational basis. The state of the system prior to channel invocation can be obtained from this by setting $\lambda =1.$ 

The presence of entanglement can be assessed by considering partial transpose of $\rhof$ and this is
\begin{equation}
	\rhof^\mathrm{PT}  = \frac{1}{4}
	        \begin{pmatrix}
	         1+\lambda^m r^2 & 0 & 0 & 0\\
					 0 & 1-\lambda^m  r^2 & 2 i r\lambda^m & 0\\
					 0 & -2 i r\lambda^m  &1-\lambda^m  r^2  & 0 \\
					 0 & 0 & 0 & 1+\lambda^m r^2
	        \end{pmatrix},
	\label{eq:finalrhofortwopt}
\end{equation}
The state is separable if $\rhof^\mathrm{PT}$ is positive~\cite{peres96,horodecki96}. The eigenvalues of this will only be positive if and only if $0\leqslant r \leqslant \sqrt{1+1/\lambda^m} -1.$ Thus the state prior to channel invocation will be separable if and only if $0\leqslant r \leqslant \sqrt{2} -1.$ The correlated state protocol can give estimation gains as $r \rightarrow 0,$ and this clearly corresponds to a state which is separable throughout the protocol. Thus the correlated state protocol gains cannot always be ascribed to the presence of entanglement. 

Quantum discord describes correlations between quantum systems in terms of mutual information and measures the extent to which these exceed any possible classical correlations between the systems~\cite{ollivier01,modi11}. It is possible to compute the quantum discord for states with the form~\cite{luo08,chen11}
\begin{equation}
	\rhoop = \frac{1}{4}\;
	              \left( 
								 \hat{I} 
								 +\sum_{j=1}^3 c_j \hat{\sigma}_j \otimes \hat{\sigma}_j
								\right)
	\label{eq:rhoforQ}
\end{equation}
where $c_j$ are all real and $ \hat{\sigma}_1 = \sigmax, \ldots. $ Defining
\begin{subequations}
 \label{eq:evalsforQ}
\begin{eqnarray}
 \lambda_0 & := & 1-c_1-c_2-c_3 \\
 \lambda_1 & := &  1-c_1+c_2+c_3  \\
 \lambda_2 & := & 1+c_1-c_2+c_3  \\
 \lambda_3 & := & 1+c_1+c_2-c_3 
\end{eqnarray}
\end{subequations}
and 
\begin{equation}
	c:= \max{ \left\{ \left| c_1\right|, \left| c_2\right|,\left| c_3\right| \right\}
	             }
\end{equation}
yields that the quantum discord is~\cite{luo08} 
\begin{eqnarray}
	{\cal Q}(\rhoop) & = & \frac{1}{4}\sum_{j=0}^3 \lambda_j \log_2{\lambda_j}  -\frac{1-c}{2}\;  \log_2{\left( 1-c \right) }  \nonumber \\
	                         & &  - \frac{1+c}{2}\;  \log_2{\left( 1+c \right) }.
\end{eqnarray}
Following the scheme of~\cite{collins13}, the final density operator can be brought into the form of Eq.~\eqref{eq:rhoforQ} by applying the single qubit unitary
\begin{equation}
 \hat{U} = 
 \begin{pmatrix}
   0 & e^{i\pi/8} \\
	  e^{-i\pi/8}  & 0
 \end{pmatrix}	
\end{equation}
to each qubit and this will not alter the discord. Then 
\begin{eqnarray}
 \lambda_0 & = & 1-\lambda^m r^2 \\
 \lambda_1 & = &  1+ 2r\lambda^m  + \lambda^mr^2 \\
 \lambda_2 & = & 1- 2r\lambda^m  + \lambda^mr^2  \\
 \lambda_3 & = & 1-\lambda^mr^2 
\end{eqnarray}
and 
\begin{equation}
	c= \max{ \left\{\lambda^m r^2, \lambda^m r \right\} = \lambda^m r
	             }
\end{equation}
since $0 \leqslant r \leqslant 1.$ Again the discord for the correlated state protocol prior to channel invocation can be determined by setting $\lambda=1$ and this gives
\begin{equation}
	{\cal Q}(\rhoop) = \frac{1+r}{2}\;
	                             \log_2{\left( 1+r \right) } 
															+ \frac{1-r}{2}\;
	                             \log_2{\left( 1-r \right) }.
\end{equation}
Then  $\frac{\partial {\cal Q}}{\partial r} >0$ whenever $r>0$ and ${\cal Q} =0$ when $r=0$ implies that the state prior to channel invocation has non-zero discord whenever the polarization of each qubit is non-zero. In this sense, we can regard the correlated state protocol as using states that have quantum correlations. 


\section{Discussion}
\label{sec:summary}

We have considered estimation of the depolarizing channel parameter for various situations where the available initial state of each qubit is mixed and we quantified the accuracy of each in terms of the quantum Fisher information after channel invocation. We showed that for certain ranges of initial state polarization and channel parameter, a protocol that uses correlated states prior to channel invocation attains a larger quantum Fisher information per channel invocation than either a protocol which uses product states throughout or another protocol which uses a sequence of channel invocations on a single qubit. For qubits with small initial polarizations we showed that using more than two qubits and more than two channel invocations can be advantageous; this is in contrast to the situation where qubits in pure initial states are available. 

Our results suggest that, as with Pauli channel parameter estimation, a single invocation of the channel on many correlated qubits gives gains for all parameter values and initial polarizations. We were only able to prove this for the depolarizing channel in the limit as the initial polarization approaches zero, but we were also unable to generate a counterexample. This was not central to our results but it does merit further investigation. 

The source of the gain cannot always be attributed to the presence of entangled or non-separable states prior to channel invocation; we demonstrated this for the two qubit example. We also showed that there is non-zero discord present prior to channel invocation in this example. A full analysis of the evolution of the discord and its possible relationship to gains in the quantum Fisher information might provide insight into the source of these accuracy gains; this is beyond the scope of our article and remains to be investigated. 

Finally, we chose one particular preparatory unitary transformation for the analysis of the correlated state protocol; that this yields gains in some cases is interesting enough. However it is unclear whether this is optimal or whether there exists another parameter independent unitary that offers greater gains.


\appendix


\section{Sequential channel use gain}
\label{app:gainseq}

The behavior of $\Gseq$ with respect to $r$ can be assessed by first setting $x:=r^2$. Then 
\begin{equation}
	\Gseq = m \frac{\lambda^{2m-2} - \lambda^{2m} x}{1 - \lambda^{2m} x}
\end{equation}
gives that
\begin{equation}
  \frac{\partial \Gseq}{ \partial x} = -m \lambda^{2m} \frac{1 + \lambda^{2m-2}}{\left( 1 - \lambda^{2m} x\right)^2} \leqslant 0.
\end{equation}
Since $r$ increases as $x$ increases this proves the fact that $\Gseq$ is a monotonically decreasing function of $r$.

To assess the behavior of $\Gseq$ with respect to $\lambda$, let $y:= 1/\lambda^2$ and note that $1 \leqslant y.$ Then 
\begin{equation}
	\Gseq = m \frac{y - r^2}{y^m - r^2}
\end{equation}
and 
\begin{eqnarray}
  y^m - r^2 & = & (y - r^2)\sum_{k=0}^{m-1} y^k r^{2m-2k-2} \nonumber \\
	                &  & + r^{2m} - r^2.
\end{eqnarray}
Thus
\begin{equation}
	\Gseq = \frac{m}{\sum_{k=0}^{m-1} y^k r^{2m-2k-2} + (r^{2m} - r^2)/(y - r^2)}.
\end{equation}
The term $(r^{2m} - r^2)/(y - r^2)$ is negative since $0 \leqslant r \leqslant 1$ and $1 \leqslant y$ and thus decreases as $y$ increases. The remaining term $\sum_{k=0}^{m-1} y^k r^{2m-2k-2}$ increases as $y$ increases. Thus $\Gseq$ decreases as $y$ increases. However, $y$ decreases as $\lambda$ increases and this proves the fact that $\Gseq$ is a monotonically increasing function of $\lambda$.

For the pure initial state case, $r=1$ and Eq.~\eqref{eq:gainseq} gives
\begin{equation}
	\Gseq = m \frac{y - 1}{y^m - 1}
\end{equation}
where $y = 1/\lambda^2$. Then 
\begin{equation}
  y^m - 1  =  (y - 1)\sum_{k=0}^{m-1} y^k 
\end{equation}
gives
\begin{equation}
	\Gseq = \frac{m}{\sum_{k=0}^{m-1} y^k }.
\end{equation}
This is a monotonically decreasing function of $y \geqslant 1$ and thus attains a maximum of $1$ at $y=1.$ Thus $0 \leqslant \Gseq \leqslant 1$ for a pure initial state, where the lower bound stems from the fact that the gain is a ration of positive quantities. 

For the mixed initial state case, $\Gseq$ is a monotonically increasing function of $\lambda$ and attains a maximum of $m$ when $\lambda=1.$ This shows that $0 \leqslant \Gseq \leqslant m$ when $r < 1.$


\section{Symmetries in the diagonal terms of $\rhof$}
\label{app:rhofdiag}

Eqs.~\eqref{eq:depoloncompbasis} give that the diagonal terms of $\rhof$ only arise from the action of the channel on 
\begin{equation}
	\rhoprepdiag := \sum_{x=0}^{(N-1)/2}
	            \diag{j(x)}\;
							\biggl[
		           \ket{x}\bra{x} + 
							 \ket{N-x}\bra{N-x}
              \biggr]
	 \label{eq:rhoprepdiag}
\end{equation}
and that this will be mapped to the diagonal component of the final density operator, denoted $\rhofdiag$ and defined via $\rhoprepdiag \stackrel{\hat{\Gamma} \otimes \cdots \otimes \hat{\Gamma}}{\mapsto}  \rhofdiag.$ A useful expression for $\rhofdiag$ involves the following representation of the channel on any diagonal operator. Eq.~\eqref{eq:depoloncompbasisdiag} gives 
\begin{equation}
	 \ket{x}\bra{x}  \stackrel{\hat{\Gamma}}{\mapsto}   p \ket{x}\bra{x} + q \sigmax\ket{x}\bra{x} \sigmax.
\end{equation}
Thus if the channel acts on the rightmost qubit, labeled $1$, then its action on the diagonal component of the density operator is
\begin{equation}
	\rhoprepdiag  \stackrel{\hat{\Gamma}^{(1)}}{\mapsto}   p \rhoprepdiag 
	                                                                      + q \xop_1 \rhoprepdiag   \xop_1 
\end{equation}
where $ \xop_1 := \iop \otimes \cdots \otimes \iop \otimes \sigmax$ and the superscript on $\hat{\Gamma}$ indicates action on qubit $1$. Suppose that the channel acts on qubit $1$ and then on qubit $2$. This maps the diagonal component of the prepared density operator as
\begin{eqnarray}
	\rhoprepdiag  & \stackrel{\hat{\Gamma}^{(1)}}{\mapsto}  &  p \rhoprepdiag 
	                                                                      + q \xop_1 \rhoprepdiag   \xop_1
																																				\nonumber \\
											& \stackrel{\hat{\Gamma}^{(2)}}{\mapsto}  &  p^2 \rhoprepdiag 
																																				\nonumber \\
	                                                & &                     + pq \bigl[ \xop_1 \rhoprepdiag   \xop_1 + \xop_2 \rhoprepdiag   \xop_2\bigr]
																																				\nonumber \\	
																							& & 	+ q^2 		\xop_2\xop_1 \rhoprepdiag 	\xop_2\xop_1																															
\end{eqnarray}
where $ \xop_2 := \iop \otimes \cdots \otimes \iop \otimes \sigmax \otimes \iop.$ This extends to action of the channel once on each of qubits $1,2,3,\ldots m$ and gives 
\begin{eqnarray}
	\rhoprepdiag  & \mapsto  &  \rhofdiag \nonumber \\
											& =  &  \sum_{k=0}^m q^k p^{m-k}  \times \nonumber \\
											& &   \sum
																							\xop_{l_k} \cdots \xop_{l_1} \rhoprepdiag 	\xop_{l_k} \cdots \xop_{l_1}				
		\label{eq:rhofinalops}
\end{eqnarray}
where the inner sum is over all $(l_k, \ldots, l_1)$ that satisfy $l_k > l_{k-1} > \cdots > l_1$ and range over the entire possibilities starting with $l_1=1$ to $l_k=m.$ If $k=0$ the inner sum is taken to mean $\rhoprepdiag$. For example, if $k=3$ and $m=4$ there are three indices $l_3>l_2>l_1$ and the summation range yields the following possibilities for $(l_3,l_2,l_1)$: $(3,2,1),(4,2,1), (4,3,1),(4,3,2).$ This consists of the set of all flips on three out of the four qubits with least significant label. A more symmetrical alternative is 
\begin{eqnarray}
	\rhofdiag      & =  &  \sum_{k=0}^m q^k p^{m-k}    \frac{1}{k!}\;  \times \nonumber \\
										 &     &  \sum \xop_{l_k} \cdots \xop_{l_1} \rhoprepdiag 	\xop_{l_k} \cdots \xop_{l_1}				
		\label{eq:rhofinalopssym}
\end{eqnarray}
where the sum is over all $m \geqslant l_k,l_{k-1}, \ldots, l_1 \geqslant 1$ where no two indices are equal. We introduce the following notation to render this more compact. Let $\vec{l}:=(l_k, \ldots, l_1)$ where, $m \geqslant l_k,l_{k-1}, \ldots, l_1 \geqslant 1$ and no two elements in this list are equal. Then define $\xop_{\vec{l}}:= \xop_{l_k} \cdots \xop_{l_1}.$ It follows that
\begin{equation}
	\rhofdiag  =  \sum_{k=0}^m     \frac{q^k p^{m-k}}{k!}\;  
										              \sum_{\vec{l}}
																							\xop_{\vec{l}}\; \rhoprepdiag \xop_{\vec{l}}\;.				
		\label{eq:rhofinalopscompact}
\end{equation}

We use this to prove that
\begin{equation}
	\bra{N-x} \rhof(\lambda)  \ket{N-x} = \bra{x} \rhof(\lambda)  \ket{x}.
	\label{eq:invarunderall}
\end{equation}
To do so, note that $\ket{N-x} = \xop_n \cdots \xop_1\ket{x}$ and let $\xop_{\textrm{tot}}:= \xop_n \cdots \xop_1.$ Thus, using $\xop_{\textrm{tot}}^\dagger = \xop_{\textrm{tot}},$
\begin{eqnarray}
	\bra{N-x} \rhof(\lambda)  \ket{N-x} & = &  \bra{N-x} \rhofdiag \ket{N-x} \nonumber \\
	                                                    & = & \bra{x} \xop_{\textrm{tot}} \rhofdiag  \xop_{\textrm{tot}} \ket{x}. 
\end{eqnarray}
Eq.~\eqref{eq:rhofinalopscompact} gives
\begin{widetext}
\begin{eqnarray}
	\bra{N-x} \rhof(\lambda)  \ket{N-x} & = &  \bra{x} 
	                                                               \sum_{k=0}^m \frac{q^k p^{m-k}}{k!}\;
																														     \sum_{\vec{l}} 
																																  \xop_{\textrm{tot}} 
																			                           	\xop_{\vec{l}}\;  \rhoprepdiag 		\xop_{\vec{l}}\;  
																			                           \xop_{\textrm{tot}} \ket{x}
																			                           \nonumber \\
	                                                    & = &  
																											          \sum_{k=0}^m  \frac{q^k p^{m-k}}{k!}\;
																														     \sum_{\vec{l}} 
																																 \bra{x} 
																			                           	\xop_{\vec{l}}\;
	                                                               \xop_{\textrm{tot}} 
																																 \rhoprepdiag 	
																																 \xop_{\textrm{tot}} 
																																 	\xop_{\vec{l}}\;	  \ket{x}
		\label{eq:totalopondiag}
\end{eqnarray}
\end{widetext}
since $\xop_{\textrm{tot}} $ commutes with $\xop_{\vec{l}}\;$. Then Eq.~\eqref{eq:rhoprepdiag} gives $\xop_{\textrm{tot}} \rhoprepdiag \xop_{\textrm{tot}}	= \rhoprepdiag.$ This and Eq.~\eqref{eq:totalopondiag} give 
%
\begin{eqnarray}
	\bra{N-x} \rhof(\lambda)  \ket{N-x} & = &  \bra{x} 
	                                                               \sum_{k=0}^m  \frac{q^k p^{m-k}}{k!}\;  \times \nonumber \\
																											&		&	     
																											           \sum_{\vec{l}} 
																			                           \xop_{\vec{l}}\;
																																 \rhoprepdiag 
																																 \xop_{\vec{l}}\;  \ket{x} \nonumber \\
																											& = &  \bra{x} \rhof(\lambda)  \ket{x}. 
\end{eqnarray}

\section{Diagonal terms of $\rhof$}
\label{app:rhofdiagterms}

Let the binary representation of $x$ be $x = x_n \ldots x_1.$ Then  $\diagf{x} = \bra{x} \rhof\ket{x}$ and Eq.~\eqref{eq:rhofinalops} give
\begin{widetext}
\begin{eqnarray}
	\diagf{x}     & =  &  \sum_{k=0}^m q^k p^{m-k}\;  
										            \sum \bra{x_n \ldots  \overline{x}_{l_k} \ldots  \overline{x}_{l_1} \ldots   x_1}
										                     \rhoprepdiag 	
																			   \ket{x_n \ldots  \overline{x}_{l_k} \ldots  \overline{x}_{l_1} \ldots   x_1}	
		\label{eq:rhofinaldiagterms}
\end{eqnarray}
\end{widetext}
with the same constraints for the inner sum as for Eq.~\eqref{eq:rhofinalops} and where $\overline{x}_l = 1-x_l$. We shall evaluate the inner sum for any value of $k$, which represents the number of flips on the bits within the binary representation of $x$. 

The fact that the diagonal terms of $\rhof$ only depend on the number of (and not their order) zeroes, $u$, in the $n-m$ most significant bits and also the number of zeroes, $v$, in the $m$ least significant bits implies that it is sufficient to consider $\diagf{x}$ where the binary representation of $x$ is
\begin{equation}
  x= \underbrace{0\ldots0}_{u} \underbrace{1\ldots1}_{n-m-u} \underbrace{1\ldots1}_{m-v} \underbrace{0\ldots0}_{v\; \textrm{terms}}.
\end{equation}
Here Eqs.~\eqref{eq:rhoprepped} and~\eqref{eq:rhofinaldiagterms} give
\begin{eqnarray}
	\diagf{x}     & =  &  \sum_{k=0}^m q^k p^{m-k}\;  
										            \sum d_{j(x_n \ldots  \overline{x}_{l_k} \ldots  \overline{x}_{l_1} \ldots   x_1)}
		\label{eq:rhofinaldiagtermstwo}
\end{eqnarray}
with the same limits for the inner sum. These limits can be modified by observing that $d_{j(x_n \ldots  \overline{x}_{l_k} \ldots  \overline{x}_{l_1} \ldots   x_1)}$ only depends on the number of zeroes and ones in the binary representation of $x$ together with the number and arrangement of flips. 
Suppose that, of the $k$ flips, $l$ occur on the rightmost cluster of $v$ bits. Thus $k-l$ occur on the second to right cluster of $m-v$ bits. Then the bit representation,  $x_n \ldots  \overline{x}_{l_k} \ldots  \overline{x}_{l_1} \ldots   x_1$, contains $u$ zeroes from the leftmost cluster, $0$ from the next cluster to the right, $(k-l)$ from the second to right cluster and $v-l$ from the rightmost cluster.  Thus the total number of zeroes in this bit string are $u+ v+ k - 2l$ and 
\begin{equation}
 d_{j(x_n \ldots  \overline{x}_{l_k} \ldots  \overline{x}_{l_1} \ldots   x_1)} = d_{u+v + k - 2l}.
\end{equation}
The exact location of the flips within each cluster is irrelevant. Thus within the inner sum of Eq.~\eqref{eq:rhofinaldiagtermstwo} there are  $\binom{v}{l} \binom{m-v}{k-l}$ ways of attaining  $d_{u+v + k - 2l}$, consisting of $\binom{v}{l}$ ways of flipping $l$ bits in the rightmost cluster together with $ \binom{m-v}{k-l}$ ways of flipping $k-l$ in the second to right cluster. 
Finally not every value of $l$ is possible and these are constrained by $0 \leqslant l \leqslant v$ (at most $v$ flips in the rightmost cluster) and $0 \leqslant k-l  \leqslant m-v$ (at most $m-v$ flips in the second to right cluster). Taken together these give $ \max{(0,v-m+k)} \leqslant l \leqslant \min{(k,v)}.$

Combining these yields Eq.~\eqref{eq:diagfinal}. 


\section{Quantum Fisher Information}
\label{app:qfifirst}

We shall show that 
\begin{eqnarray}
  H_x & = &  \frac{2}{\diagf{x}^2 - \lambda^{2m} \cdiag{j(x)}^2 } \times 
		                 \left[ \diagf{x} \left(  \dotdiagf{x}^{\phantom{f}2} + m^2 \lambda^{2m-2} \cdiag{j(x)}^2 \right) \right.   \nonumber \\
     &   &       \left. - 2 m\lambda^{2m-1} \dotdiagf{x} \cdiag{j(x)}^2
		          \right]
\end{eqnarray}
and the result will follow from Eq.~\eqref{eq:qfisum}. The expression for $H_x$ comes from the eigenvalue decomposition of $\rhofx.$ The eigenvalues are of $\rhofx$
\begin{equation}
     p_\pm := \diagf{x} \pm \lambda^m \cdiag{j(x)}
\label{eq:rhofevals}
\end{equation}
and the associated eigenvectors are
\begin{equation}
     \ket{\phi_\pm} := \left(\ket{x} \mp i \ket{N-x} \right) /\sqrt{2}.
\label{eq:rhofevecs}
\end{equation}
The fact that the eigenvalues are independent of $\lambda$ means that the second term on the right hand side of Eq.~\eqref{eq:qfieigenval} is zero and the first term then yields
\begin{eqnarray}
  H_x & = & \frac{1}{\diagf{x} + \lambda^m \cdiag{j(x)}}\;
	                \left( \dotdiagf{x} + m\lambda^{m-1} \cdiag{j(x)} \right)^2
									\nonumber \\
				& &  + \frac{1}{\diagf{x} - \lambda^m \cdiag{j(x)}}\;
	                \left( \dotdiagf{x} - m\lambda^{m-1} \cdiag{j(x)} \right)^2.
\label{eq:Hxstepone}
\end{eqnarray}
Straightforward algebra yields the desired result.

The expression for the quantum Fisher information can be rewritten as a sum over $u$ and $v$. The only relevance that each value of $x$ has for the summand is that it determines  $u, v$ and $j(x) = u+v.$ Typically many values of $x$ will yield the same combination of values of $u$ and $v$. To account for this, note that the sum in Eq.~\eqref{eq:qfidepolone} runs from $x = 00 \ldots 0$ to $x = 01 \ldots 1$  (in binary notation). The fact that the leftmost bit is always zero gives two distinct cases. 

In the first, where $m<n,$ this bit structure reveals that $1\leqslant u \leqslant n-m$ and $0\leqslant v \leqslant m$. For any choice of $u$, there are $\binom{n-m-1}{u-1}$ ways in which the zeroes can appear in the leftmost $n-m$ bits, given that the leftmost single bit must be zero. There are independently $\binom{m}{v}$ ways in which the zeroes can appear in the rightmost $m$ bits. Thus there are $\binom{n-m-1}{u-1}\binom{m}{v}$ values of $x$ in the range of the sum which result in exactly the same pair of values $u$ and $v$. For each the contribution of the summand is identical. This means that the terms within the summand can be expressed in terms of $u$ and $v$ (rather than $x$) and the result sum reduces to that of Eq.~\eqref{eq:qfidepoltwomlessthann}.

In the second, where $m=n$, clearly $u=0$ and $ 1 \leqslant v \leqslant n-1.$ A similar counting argument reveals that there are $\binom{v-1}{n-1}$ ways in which the zeroes can appear in the rightmost $n-1$ bits, since the rightmost bit must be zero. A similar argument then results in Eq.~\eqref{eq:qfidepoltwomequalsn}.


\section{Low polarization correlated state protocol quantum Fisher information}
\label{app:smallrqfi}

The correlated state protocol quantum Fisher information, Eqs.~\eqref{eq:qfidepoltwomlessthann} and~\eqref{eq:qfidepoltwomequalsn}, requires $\cdiag{j}$ and $\diagf{}(u,v),$ which in turn requires $\diag{j}.$ Eqs,~\eqref{eq:prepcounterdiag} and~\eqref{eq:prepdiag} yield, to second order in $r$, 
\begin{eqnarray}
 \cdiag{j} &\approx & \frac{(2j-n)}{2^{n}}\, r \quad \textrm{and} \label{eq:cdiagapprox} \\
 \diag{j} &\approx & \frac{1}{2^n} + \frac{r^2 \left[ ( 2j -n) ^2-n \right]}{2^{n+1}}.
\end{eqnarray}
The latter of these gives
\begin{equation}
  \diagf{}(u,v)= \alpha + \beta r^2
	\label{eq:diagfinalapprox}
\end{equation}
where
\begin{equation}
	\alpha := \sum_{k=0}^{m} q^k p^{m-k} 
	                                  \sum^{l_\mathrm{max}}_{l=l_\mathrm{min}}\;
																		\binom{v}{l}
																		\binom{m-v}{k-l}\;  \frac{1}{2^n} 
\end{equation}
and
\begin{eqnarray}
	\beta & := & \sum_{k=0}^{m} q^k p^{m-k} 
	                                  \sum^{l_\mathrm{max}}_{l=l_\mathrm{min}}\;
																		\binom{v}{l}
																		\binom{m-v}{k-l}\; \nonumber \\
					& & 					\times\frac{\left[ ( 2u+2v+2k-4l -n) ^2-n \right]}{2^{n+1}}.
\end{eqnarray}
Both $\alpha$ and $\beta$ are independent of $r$. Then
\begin{equation}
	\alpha =   \frac{1}{2^n}, 
	\label{eq:alpha}
\end{equation}
which can be demonstrated indirectly by noting that, when $r=0,$ each diagonal term of the final density operator equal $\alpha.$ However, when $r=0,$ $\rhoo= \iop/2$  and thus $\rhoprep = \iop/2^n$. The channel does not alter this and thus, when $r=0$, $\rhof = \iop/2^n$. It follows that $\alpha = 1/2^n.$ This fact can independently be verified by explicitly evaluating the three summations in $\alpha.$ 

Note that these imply
\begin{equation}
 \dotdiagf{}^{2}  \approx \frac{\partial \beta}{\partial \lambda}\; r^2.
\end{equation}
Then Eqs~\eqref{eq:cdiagapprox},~\eqref{eq:diagfinalapprox} and~\eqref{eq:alpha}, yield that for the terms in Eqs.~\eqref{eq:qfidepoltwomlessthann} and~\eqref{eq:qfidepoltwomequalsn}
\begin{eqnarray}
	\diagf{}^2 - \lambda^{2m} \cdiag{u+v}^2  &\approx &   \frac{1}{2^{2n}}, \nonumber \\
	\diagf{} \left(  \dotdiagf{}^{2} + m^2 \lambda^{2m-2} \cdiag{u+v}^2 \right) & \approx & \diagf{} m^2 \lambda^{2m-2} \cdiag{u+v}^2 \nonumber \\ 
	                                                                                                                 & = & \frac{m^2 \lambda^{2m-2}}{2^{3n}} \left( 2u + 2v-n\right) r^2 \nonumber \\
	\dotdiagf{}^{2}\cdiag{u+v} &\approx & \left( \frac{\partial \beta}{\partial \lambda} \frac{2(2u + 2v -n)}{2^{n+1}}\right)^2\, r^6. \nonumber
\end{eqnarray}
Thus if $m<n$, Eqs.~\eqref{eq:qfidepoltwomlessthann} becomes
\begin{eqnarray}
 \Hcorr & \approx & \sum_{v=0}^{m} \sum_{u=1}^{n-m} \binom{n-m-1}{u-1} \binom{m}{v}\;
              2 ^{2n+1} \nonumber \\
     &   & \! \! \!
		          \times  \frac{m^2 \lambda^{2m-2}}{2^{3n}} \left( 2u + 2v-n\right) r^2 \nonumber \\
			& = & r^2\; \frac{m^2 \lambda^{2m-2}}{2^{n-1}} \sum_{u=1}^{n-m} \binom{n-m-1}{u-1}  \nonumber \\
			&  & \times   \sum_{v=0}^{m} \binom{m}{v} \left( 2u + 2v-n\right).
\label{eq:qfidepoltwomlessthannapprox}
\end{eqnarray}
The two summations in Eq~\eqref{eq:qfidepoltwomlessthannapprox} give $n 2^{n-1}$ and this yields the result of Eq.~\eqref{eq:qfiseqperapprox}. A similar derivation gives the same result when $m=n.$


%

\end{document}